\documentclass[conference]{IEEEtran}
\IEEEoverridecommandlockouts
\usepackage{cite}
\usepackage{amsmath,amssymb,amsfonts}
\usepackage{graphicx}
\usepackage{multirow}
\usepackage{textcomp}
\usepackage{xcolor}
\usepackage[caption=false]{subfig}
\usepackage[]{algpseudocode,algorithm}

\def\BibTeX{{\rm B\kern-.05em{\sc i\kern-.025em b}\kern-.08em
    T\kern-.1667em\lower.7ex\hbox{E}\kern-.125emX}}
\begin{document}
\title{Botnet Detection Using Recurrent Variational Autoencoder\\
}

\author{\IEEEauthorblockN{Jeeyung Kim\IEEEauthorrefmark{1},
Alex Sim\IEEEauthorrefmark{1}, Jinoh Kim\IEEEauthorrefmark{2}, 
Kesheng Wu\IEEEauthorrefmark{1}}
\IEEEauthorblockA{\IEEEauthorrefmark{1}Lawrence Berkeley National Laboratory, 
Berkeley, CA, USA\\
\IEEEauthorrefmark{2}Texas A\&M University, Commerce, TX, USA\\
Email: \IEEEauthorrefmark{1}\{jeeyungkim, asim, kwu\}@lbl.gov,
\IEEEauthorrefmark{2}jinoh.kim@tamuc.edu}}

\maketitle
\begin{abstract}
Botnets are increasingly used by malicious actors, creating increasing
threat to a large number of internet users.
To address this growing danger, we propose to study methods to detect
botnets, especially those that are hard to capture with the commonly
used methods, such as the signature based ones and the existing anomaly-based ones.
More specifically, we propose a novel machine learning based method, named Recurrent Variational Autoencoder (RVAE), for detecting botnets through sequential characteristics of network traffic flow data including attacks by botnets.

We validate robustness of our method with the CTU-13 dataset, where we have chosen the testing dataset to have different types of botnets than those of training dataset.
Tests show that RVAE is able to detect botnets with the same accuracy as the best known results published in literature.
In addition, we propose an approach to assign anomaly score based on
probability distributions, which allows us to detect botnets in streaming
mode as the new networking statistics becomes available.
This on-line detection capability would enable real-time detection of
unknown botnets.
\end{abstract}


\section{Introduction} 
Botnet is one of the most significant threats to the cyber-security as they are considered a source of many malicious activities~\cite{alieyan2017survey}.
The machines in a botnet is typically hijacked without the owner's
knowledge.
These machines are then commanded to act together to attack more
machines and find more  valuable targets.
They are also frequently used to perform
distributed denial-of-service attacks (DDos), click-fraud, spamming and crypto-mining.
These botnets could  also harbor malware and ransomware for delivery to
victims of their attacks.
Therefore, a critical task of cybersecurity  research is to detect botnets and stop their attacks.

At the same time, the malicious software for infecting a machine and
operating botnets is evolving to evade detection, rendering many of the
commonly used botnet detection techniques ineffective. 
For example, the protocol used by botnet has changed.
Initially, Internet Relay Chat (IRC) in which the bot master controls the other bots was adopted as the communication method.
After that, peer-to-peer (P2P) where individual boats serve as clients and servers was widely used and then HTTP based botnet hijacking a legitimate communication channels began to flourish~\cite{zhao2013botnet}.
Moreover, botnet attack techniques are continuously evolving.
In 2016, a new botnet named Mirai started to control hundreds of thousands of Internet of Things (IoT) devices making a high-profile DDoS threat. Mirai has made other botnets that imitate the infection strategy~\cite{antonakakis2017understanding}.
The other sophisticated botnet systems, such as Smominru which is known as crypto-mining botnet became a rampant threat since 2018.
There are also examples where botnets could operate without detection
for an extended period of time by gradually changing its mode of
operation.
Detecting these botnets requires the detection algorithms to
evolve with time and to adapt quickly with extensive retraining.

There is a significant body of published literature on detecting malicious botnets.
The existing approaches of botnet detection are categorized into two
broad categories: honeypots and Intrusion Detection Systems (IDS).
Honeypot is a computer mechanism that is used as trap to draw the attention of attackers to attack this computer system~\cite{zeidanloo2010taxonomy}.
The honeypot approach has several limitations in terms of scale and possible detection types of attacks~\cite{zeidanloo2010taxonomy}.
On the other hand, IDS methods which is to monitor a network or systems for malicious activity are further divided into two categories: signature-based and anomaly-based methods.
A signature-based method is configured with a set of rules or signatures to classify types of network traffics.
This approach requires a relatively small amount of computation and
often can work in real-time without slowing down normal network operations.
The effectiveness of signature-based is widely studied, but it is only able to be identify well-known botnets~\cite{zeidanloo2010taxonomy, roesch1999snort}.

On the other hand, anomaly-based  techniques detect botnets based on a number of network traffic anomalies such as high network latency, high volumes of traffic and unusual system behavior~\cite{zeidanloo2010taxonomy}.
They model a normal behavior of network traffic and design a decision engine which determines any divergence or statistical deviations from the norm as a threat~\cite{abou2020botchase}.
Traditionally, many studies have attempted to use statistical features or heuristic methods to detect botnet anomalies~\cite{binkley2006algorithm,gu2008botsniffer}.
Recently, with the motivation of making more generalized botnet detectors in anomaly-based IDS, there have been many studies on machine learning (ML) methods to analyze botnets behavior for the anomaly detection. 
With anomaly detection system utilizing ML, previously unseen types of botnet attacks can be detected based on its behavior~\cite{venkatesh2012http, singh2014big, beigi2014towards, stevanovic2014efficient, zhao2013botnet}.

However, many studies suggesting ML methods for botnet detection are limited in that they do not share the testing dataset, which makes them incomparable to each other.~\cite{venkatesh2012http, singh2014big, beigi2014towards, stevanovic2014efficient, zhao2013botnet}.
In addition, most previous studies do not take into account sequential patterns within network traffic data, even though botnet traffic shows a repeated pattern behavior due to the nature of the pre-programmed characteristics of bots~\cite{assadhan2009detecting}.
There are some studies considering periodic behaviors, but they are still limited as the studies only consider sequential characteristics within the same source IP addresses not within overall network traffic, which makes it difficult to be used as the online detection system~\cite{sinha2019tracking, torres2016analysis}.
Furthermore, some existing studies narrow their scope by evaluating their methods only for one of the IRC, P2P, and HTTP traffic.
For these reasons, those methods fail to validate their methods to cope with various types of botnets or previously unseen botnet families, limiting the practical use given the new introduction of new botnet families~\cite{nguyen2019gee, nicolau2018learning, sinha2019tracking,torres2016analysis}. 
Only the method which shows effectiveness on various types of botnets can be rendered reliable and practically useful.

Our goal of this paper is to suggest a ML method which is capable to reflect periodicity within network data as well as to detect previously unseen types of botnets in a on-line manner. We have three main contributions in this paper:
\begin{itemize}
\item We adapt Recurrent Variational Autoencoder (RVAE) architecture for
anomaly detection. This detection method can be trained on normal
data and detect anomalies that vary over time. This is a novel
feature of our detection approach.
\item We devise a strategy for on-line detection of anomalies using the
output from the RVAE network.
\item We verify that the on-line detection approach could detect changing
botnets by splitting the popular test data set CTU-13 into training
and testing sets with different types of botnets. Tests show that we
are able to detect botnets effectively when
different types of botnets are used for testing compared to the existing methods.
\end{itemize}

\section{Related works and Background}
The various ML method have been utilized in botnet detection. We use a fundamental structure of Recurrent VAE (RVAE) which contains both Variational Autoencoder (VAE) and Recurrent Neural Network (RNN) in regard that it uses RNN structure as encoder/decoder of VAE instead of  Multi-Layer Perceptron (MLP). 
In the next subsection~\ref{mlap}, we introduce previous work utilizing ML techniques for botnet detection and discuss the limitation each work has. Subsequently, in the subsection~\ref{mlst}, regardless of the topic of botnet detection, we describe each part of the proposed ML model, how each model works, and what problems each method was created to deal with. We also explain why the method need to be utilized for botnet detection.  

\subsection{Machine Learning Approach for Anomaly based IDS}\label{mlap}
\textbf{Variational Autoencoder:}
In ~\cite{nguyen2019gee}, Guoc et al. introduce VAE which is an unsupervised method for detecting anomalies and also focus on explaining anomalies with a gradient-based fingerprinting technique, but it is limited it assumes that they already know the ratio of anomaly and it does not consider sequential pattern of data which can increase the performance of detection.
Van et al. propose the revised VAE structure, called as Dirac Delta VAE, for achieving better anomaly detection performance in ~\cite{nicolau2018learning}. It narrows down the range of latent space which makes classifiers detect anomaly easier. However, the proposed method in the paper cannot be trained end-to-end because it separately uses classifier in the latent space. Furthermore, the authors conducted experiments using one type of botnets for training and testing. 
While it is not VAE, Ruggiero et al. utilize Denoising Autoencoder (DAE) for botnets anomaly detection in ~\cite{dargenio2018exploring}. The authors propose the way using both DAE and filter-encoder architecture to extract features for botnets classifier. 
Furthermore, there are various of studies utilizing VAE for anomaly detection of network traffic, as you can see in ~\cite{an2015variational} and ~\cite{kumagai2019transfer}.
While many works have been done so far, most are limited in that the methods overlook sequential characteristics within network traffic.

\textbf{Recurrent Neural Network}:
RNN have been in great use in many studies for employing sequential characteristics of network traffic data. Kapil et al. propose supervised approach to detect botnet hosts by tracking a host's network activity over time using RNN architecture and extract graph-based features of NetFlow data for botnet detection in ~\cite{sinha2019tracking}. However, extracted features are obtained each host IP address. If using periodicity of each source IP address, detecting malicious botnets in an timely manner is impossible because we need to wait to collect every flow with the IP address to classify one connection as {\em malicious} or {\em non-malicious}. In addition, the method is restricted in terms of not being generalized in that it is trained and tested on limited botnet scenarios.
Besides, Pablo et al. assign the symbol to size and the port and embedded the code to distributed representation like word embedding~\cite{torres2016analysis}. It shows the potential to use RNN as botnet detection model, but it is limited that it doesn't show comparable performance in imbalanced network traffic. 
In ~\cite{ongun2019designing}, Egon et al. present a method of determining which alerts are correlated by applying Neural Networks and clustering. It utilizes text strings output from IDS as RNN input. It is somehow limited as the method requires the specific output, such as text strings. 
While many works have been done so far, most are limited now that the method cannot be applied to the online anomaly detection system because the methods analyze traffic by the same host.

\textbf{Other Machine Learning Approach}:
Besides VAE and RNN, many recent studies have attempted to make use of various Machine Learning (ML) approach to reduce the dependence for human heuristics.
In ~\cite{du2019fenet}, the authors regard every feature as sentence and embed it. With embedded features, classifier can be trained to detect malicious botnets. 
In ~\cite{venkatesh2012http}, the authors propose the way of detecting HTTP botnets using MLP.
Kamaldeep et al. introduce the framework for P2P botnet detection using Random Forest in ~\cite{singh2014big}.
In ~\cite{beigi2014towards}, Elaheh et al. introduce the method of selecting effective features for machine learning based botnet detection approaches. The authors also assess its effectiveness on the dataset which are constructed focusing on generality, realism and representativeness. 
Ongun et al. present how to extract features which are good for ML model in ~\cite{ongun2019designing}. The authors also compare a statistics aggregated feature processing method with the connection level feature processing method and validate those methods with Random Forest and Gradient Boosting, which are ML techniques.
In ~\cite{taheri2018leveraging}, the authors treat network traffic features as an image. By bringing pre-trained Convolutional Neural Network (CNN) model which is suitable for image data, the authors do transfer learning to adapt network traffic data.

\subsection{Background of ML structures Employed in the Proposed Model}\label{mlst}

\textbf{Variational Autoencoder (VAE)}
VAE is one of the generative models which utilizes deep neural network structure to represent transformation.
Encoder which consists of neural network extracts latent variable $z$ in accordance with input $x$ using a reparameterization trick. Decoder which also consists of neural network reconstructs $x$ with $z$ that is created by the encoder. 
The more detailed discussion of VAE can be referred in ~\cite{kingma2013auto}.

\textbf{Gated Recurrent Unit (GRU)}
We use RNN structure which is known as containing directed cycle beneficial to represent data with sequential pattern. Especially, we utilize GRU model which has capability to remember values with long sequences comparing to vanilla RNN.
This allows GRU to extract long-term periodic characteristics of network traffic data. 
The more detailed discussion of GRU can be referred in ~\cite{chung2014empirical}.

\textbf{Recurrent Variational Autoencoder (RVAE)}
RVAE is the structure of combining seq2seq with VAE, whose encoder and decoder consists of auto-regressive model.
As it utilizes RNN instead of MLP or CNN to generate sequential outputs, it not only takes the current input into account while generating but also its neighborhood. 
For prior distribution, it uses Gaussian distribution like VAE. 
The last hidden state is used as mean and variance of multivariate Gaussian in latent space. The latent variable is employed as initial hidden state of the decoder which is also RNN structure.
The more detailed discussion of Recurrent VAE can be referred in ~\cite{bowman2015generating} and ~\cite{roberts2018hierarchical}.

Generally, seq2seq models are actively being used in text and music field. With sequential patterns, they generate music or text sequences. The advantage to the RVAE is that it utilizes sequential patterns to generate the data. With this structure, we expect that the structure performs ably in network traffic data to detect anomaly.

\section{Proposed Model}

\begin{figure}[!tb]
    \centering
    \includegraphics[width=0.9\columnwidth]{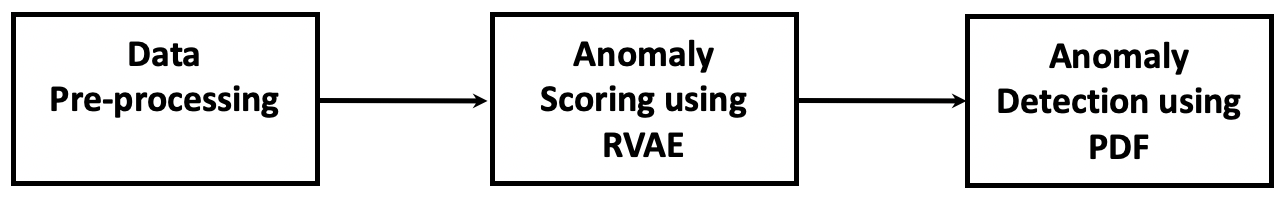}
    \caption{Procedure of the proposed method}
    \label{fig:proc}
\end{figure}

In order to identify botnets, we propose a novel flow-based botnet detection system coping with periodicity of traffic flow. 
The overall procedure of our proposed system is shown in Fig.~\ref{fig:proc}, which consists of three steps as follows:

\begin{itemize}
    \item {\bf Data pre-processing:} 
    The data instances are grouped based on a predefined time interval(e.g., 60 sec for the size of time window), and they are aggregated by host IP addresses. This process also includes the process of calculating statistical features and normalizing numerical values.
    \item {\bf Anomaly scoring:}
    At every time window, anomaly scores of every flow are calculated, which provides the degree of maliciousness of individual connections. 
    For scoring, we establish a function that consists of RVAE and produces anomaly scores by comparing the input with the output of the model which is the reconstructed input. 
    \item {\bf Anomaly detection:}
    Based on the calculated anomaly scores, the anomaly detection function classifies individual connections into either {\em Malicious} or {\em Non-malicious}.
    In particular, our method does not rely on {\em threshold}; rather, it utilizes a couple of probability density function (PDF) which are estimated by normal and botnet instances in training dataset, respectively.
\end{itemize}

\begin{figure*}[]
    \centering
    \includegraphics[width=0.85\textwidth]{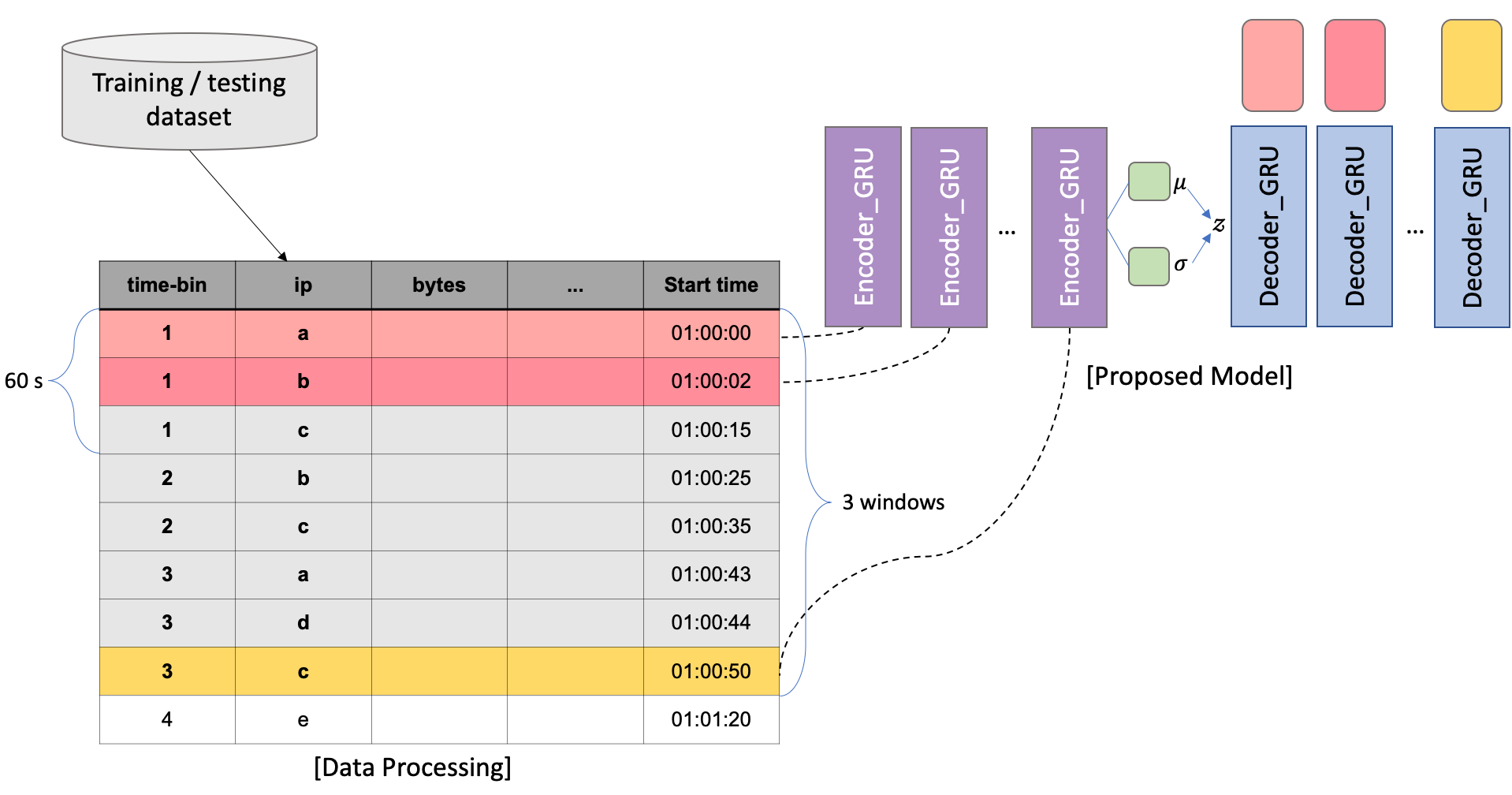}
    \caption{Botnet detection system using RVAE with sequential dataset}
    \label{fig:model structure}
\end{figure*}

Fig.~\ref{fig:model structure} demonstrates a snapshot of the process for the data pre-processing and anomaly scoring. In the phase of data processing, every flow sorted in chronological order is aggregated to obtain statistic features within the windows. These flow-based features are used as input to RVAE, and are input in the order of time. In the botnet detection system, the encoder is expected that it is trained in a way of distilling the common characteristics within the sequential data into latent variable $z$. The decoder reconstruct sequential inputs utilizing $z$. 
In the end, reconstruction loss is obtained as an output of the process of anomaly scoring.

\subsection{Anomaly Scores from Recurrent Variational Autoencoder}
\begin{table}[htbp]
\caption{Notations used}
\label{tbnotation}
\begin{center}
\begin{tabular}{|l|l|}
\hline
Notation & Description\\
\hline

$h_{E,T}$  & The last hidden state of the encoder\\
$W_\mu$ & Linear transformation to get $\mu(x)$ \\
$W_\sigma$ & Linear transformation to get $\sigma(x)$ \\
$z$ & The latent variable\\
$h_{D,2}$ & The second hidden state of the decoder\\
$h_{D,t}$ & The hidden state of the decoder at timestep $t$\\
$W^{hh}$ & Linear transformation from the previous hidden state\\
$W^{hx}$ & Linear transformation from input \\
$W^{s}$ & Linear transformation from hidden state to get output \\
$\theta$ & The parameters of encoder \\
$x_1$  & The first input of the decoder\\
$\phi$ &The parameters of decoder\\
$\tilde{y_t}$ & Output, reconstruction at timestep $t$\\
$\tilde{y_{t_n}}$ & $n$th feature of reconstruction at timestep $t$\\
$D_{KL}$ & Kullback-Leibler divergence\\
\hline
\end{tabular}
\end{center}
\end{table}
Notations used in this paper are in Table~\ref{tbnotation}.
We first input network traffic data, which are pre-processed, to GRU structure. 
$h_{E,T}$ is used as mean and variance of Gaussian distribution which represents latent space.
With $\mu$ and $\sigma$, $z$ can be obtained, and the $z$ is used as initial hidden state for the decoder. 
\begin{equation}
\begin{gathered}
    \mu(x) = W_\mu h_{E,T}\\
    \sigma(x) = W_\sigma h_{E,T}\\
    z = \mu(x) + \sigma(x)* \epsilon, \epsilon \sim N(0,1)
\end{gathered}
\end{equation}

The second hidden state of decoder follows as:
\begin{equation}
     h_{D,2} = sigmoid(W^{hh}z + W^{hx}x_{1})\\
\end{equation}
The first input of the decoder($x_1$) is zero-padded.
Finally, the outputs we obtain from RVAE is formulated:
\begin{equation}
     \tilde{y_t} = sigmoid(W^{s}h_{D,t})
\end{equation}
The loss function that we want to minimize:
\begin{equation}
    J(x) = - \mathbb{E}_{q_\phi(z|x)}[log p_\theta(x|z)] + \beta * D_{KL}[q_\phi(z|x)|p_\theta(z)]
\end{equation}
We train the model with only {\em non-malicious} instances, and in evaluation phase, we calculate reconstruction errors and use it as anomaly scores using both {\em non-malicious} and {\em malicious} instances. 
As we use binary cross entropy as error function, the anomaly score is formulated:
\begin{equation}
    L = \sum_{n=1}^{N} {(1-y_{t_n}) log (1-\tilde{{y}}_{t_n}) + y_{t_n} log \tilde{{y}}_{t_n}}
\end{equation} 
Each time window, we can obtain the anomaly scores of every connection which belong to the time window. 
In other words, if the traffic connection can be considered malicious or not is indicated by the outputs($L$) of the anomaly detection system. In the following section, we present how to detect botnets with anomaly scores.

\subsection{Anomaly Detection}\label{da}
In many studies, threshold of anomaly scores is used to distinguish whether the source IP addresses in the time window is malicious or not in anomaly detection methods~\cite{nguyen2019gee, dargenio2018exploring, an2015variational}. The threshold can be set in many ways. It is one of the simple and intuitive method; however, the information of the dataset is required in most cases such as the ratio of botnets or at least approximate values of anomaly scores of botnet samples. 
Unfortunately, there are few cases that the information about the traffic data is known in advance. 
Furthermore, detecting attacks of botnet with threshold hampers the anomaly detection framework performing on-line.
Let's say that the threshold is set to 10\%, which means that samples higher than top 10\% of anomaly scores are classified as botnets.
Then, we have to wait until all samples in testing dataset complete to get the anomaly scores because we must sort every anomaly scores. 
Therefore, this method is limited to being used in timely manner, which means that it is not practical. 

Instead, we suggest a more efficient method using the estimated probability distribution of reconstruction errors. 
In training phase, we collect reconstruction errors from normal and abnormal instances. Then, we find the distribution and its parameters to represent the distribution of reconstruction errors of abnormal and normal, respectively, by exploring various types of distributions and selecting the minimum sum of squared estimate errors (SSE). We call the distribution with the smallest SSE as the {\em best-fit PDF}. We search for the {\em best-fit PDF} among many different candidates such as {\em gamma} distribution, {\em generalized} {\em logistic} distribution, {\em fold} {\em cauchy} distribution, {\em Mielke} distribution and {\em beta} distribution, among others. In testing phase, the estimated PDF can be utilized to obtain likelihood to belong to each distribution. Comparing the likelihood values of the two different distributions, we assume that each sample of the test data set belongs to the distribution with greater likelihood. 
Utilizing {\em best-fit PDFs} at the training stage does not require the information of network traffic dataset as well as provide the botnet detection system that can be used in on-line.


\section{Experiments}\label{AA}
We have experimented several ways to validate reliability of the proposed method in different aspects. We show the two aspects from the experiments.
The first is to show that the proposed method has better performance than both Random Forest and the existing standard VAE, which we call as MLP-VAE in this paper, in various measures.
Second, we explain how the reconstruction errors are distributed and how to utilize it in detecting botnets.

\subsection{Evaluation Datasets}\label{B}
We use CTU-13 dataset which is widely used in the latest studies for botnet detection~\cite{ nguyen2019gee,  nicolau2018learning, dargenio2018exploring, an2015variational, torres2016analysis, du2019fenet, ongun2019designing, garcia2014empirical}.
A botnet scenario is a particular infection of the virtual machines using a specific malware. Thirteen of these scenarios were created, and each of them was designed to be representative of some malware behavior~\cite{garcia2014empirical}.
To compare the results of MLP-VAE and Random Forest, we reproduced nearly the same experimental settings with the settings in~\cite{nguyen2019gee} and ~\cite{ongun2019designing}.
In~\cite{nguyen2019gee} and~\cite{ongun2019designing} which proposes VAE and Random Forest structures respectively that we select as the baseline, they prove the robustness of their methods on scenario 1, 2 and 9 of CTU-13 dataset, which consists of only Neris botnet. The Neris botnet is IRC based bot infecting other machines by Spam and Click Fraud. In our reproduced experiments, all methods show similar performance in every metrics, as you see in Table~\ref{table4}. Especially, Random Forest performs very well on the testing datasets because botnet families in the testing dataset are already used for training. In other words, Random Forest method is able to capture dominant features to classify anomalies. However, when considering the evolving botnets, the method cannot be validated with being evaluated on dataset consisting of botnets which are previously identified.

Thus, we determine to follow the dataset separation criteria, as suggested in ~\cite{garcia2014empirical}, in order to test the model in more general cases.
In~\cite{garcia2014empirical}, the authors made the dataset in a way that none of the botnet families used in the training and cross-validation datasets should be used in the testing dataset. 
The authors state that this way ensures that the detection methods can generalize and detect new behaviors. 
By splitting of CTU-13 data in the suggested way, we can mimic the real situation where the operations of botnet changes over time in terms of protocols and attack types. 
Compared to the restricted dataset (scenario 1, 2, 9), various types of botnets that have IRC-based, P2P-based and HTTP-based communication methods and conduct attacks such as Spam, Click Fraud, Port Scan, DDos and FastFlux are included in the dataset we use for the experiments.
The description of dataset is in Table~\ref{table3}.

\begin{table}[htbp]
\caption{CTU-13 Dataset}
\label{table3}
\begin{center}
\begin{tabular}{|c|c|c|c|c|}
\hline
Dataset&Scenario\\
\hline
Training\&Validation&3,4,5,7,10,11,12,13\\
Testing &1,2,6,8,9\\
\hline
\end{tabular}
\end{center}
\end{table}

\subsection{Data Pre-processing}\label{AA}
The CTU-13 dataset consists of NetFlow files which are composed of source and destination IP addresses and ports, time, protocol, duration, number of packets, number of bytes, state, and service. We process the data to use the aggregated flows statistic, which is the way many existing works adopt in order to obtain flow-based features~\cite{nguyen2019gee, ongun2019designing, nicolau2018learning, dargenio2018exploring, an2015variational, torres2016analysis}. 
We group NetFlow data at every time interval of $T$, and aggregate features within every group based on the source IP addresses to get flow-based features. 
With the processing method, we can detect IP address showing malicious behavior in a particular time window. 
Many existing works experimentally find the most appropriate time window $T$, which is crucial in that while too small time window might not capture traffic characteristics over a longer period of time, too large time window cannot provide timely detection in waiting the end of the window~\cite{zhao2013botnet, nguyen2019gee, garcia2014empirical, ongun2019designing, dargenio2018exploring, an2015variational, torres2016analysis}.
We did experiment as changing the duration of windows to find the ideal value for the statistical aggregation.
We then sort the entire data within the time window by the time of the source IP connection group, because the RNN model is sensitive to the order of the inputs. 
For RVAE, we use the network traffic connections collected within $N$ windows as the sequential inputs to the model. 
You can see it in the data processing part of Fig.~\ref{fig:model structure}. 
In the case of Fig.~\ref{fig:model structure}, 60-second duration of three windows are used. 

In terms of source/destination ports and destination IP addresses, we count the number of unique records with connected source IP addresses in the time window. 
In addition, for the source IP addresses, we count the number of connections with the source IP addresses in the time window. 
For service, state, and protocol, we count the number of different values in each category with the source IP addresses in the time window. 
Finally, we normalize the numerical values to be between 0 and 1. 
As a result, the number of features used in this experiment is smaller compared to the number of features used in~\cite{ongun2019designing} and~\cite{nguyen2019gee}.

\subsection{Experimental Setting}
For splitting datasets for training, validation and testing, we followed the suggested separation criteria, as we mentioned in the section~\ref{B}. 
The architecture of MLP-VAE follows what is used in~\cite{nguyen2019gee}, [\# of features $\to$ 512 $\to$ 512 $\to$ 1024 $\to$ 100].
For RNN architecture, we use 2-layer bidirectional GRU. We use the 512 dimensions of hidden states, and 100 dimensions of latent variable as MLP-VAE. 
We also apply {\em ReLU} activation~\cite{nair2010rectified} to MLP-VAE as well as RVAE. 
The Kullback-Leibler annealing method is set so that the weight multiplied to KLD increases linearly for 500 gradient updates for RVAE. 
We train for 500 epochs with Adam optimizer and 128 batch-size.
Also, learning rate is set as 0.01 for both VAE models.
We use the 5 different evaluation metrics to validate our performance; Area Under the Receiver Operating Characteristics (AUROC), Area Under the Precision-Recall Curve (AUPRC), Precision, Recall and F1 score which are common metrics for anomaly-based IDS.
We save the model showing the best value of AUPRC in 5-fold cross validation sets.
The source code is written with the PyTorch\footnote{https://pytorch.org} library. 

\subsection{Baseline}
We compared our experimental results to the reproduced results from the MLP-VAE and Random Forest. In terms of MLP-VAE, the experimental results are based on the same data processing method with the same optimizer, learning rate and the size of latent variable from our experiment. 

\section{Results and Discussion}

We did experiments to validate the proposed method in different aspects. For quantitative validation, we compare the performance of our proposed method with other methods on different metrics. 
For qualitative validation, we plot the distribution of the reconstruction errors of normal and botnets cases. Moreover, we plot estimated the {\em best-fit PDF}.
We validate our detection method utilizing the {\em best-fit PDF} as we describe it in the section~\ref{da}. 
To compare methods in various metrics with the same processing  and detection method, we reproduce MLP-VAE and Random Forest in~\cite{nguyen2019gee} and~\cite{ongun2019designing}, respectively.  
While the value in the literature~\cite{nguyen2019gee} is 0.936, our reproduced value of AUROC with MLP-VAE is 0.966. 
Even if there may be a slightly different experimental setup, the reproduced results show that our implementation is still valid according to a bit higher result than the result of the original paper.

\textbf{Results comparison among methods} 
In Table~\ref{table4}, Random Forest method shows the nearly perfect performance in every metric, even though VAE models show the comparable performance. 
It is because that the training/testing dataset which are based on scenario 1, 2, 9 share the same characteristics. 
Random Forest is effective in finding dominant features in these characterized datasets. 
However, as we mention in the section~\ref{B}, validating the models on the characterized datasets is not what we focus on in this paper. 

In Table~\ref{table5}, we show the results from the training and testing on the generalized dataset that we mentioned in the section ~\ref{B}. In this experiment, we pre-processed our data by using 60-second duration of window and using three windows.
While both training and testing datasets that we use in Table~\ref{table4} consist of only Neris botnet, the testing dataset and training dataset we use in Table~\ref{table5} consist of each different botnets: Rbot, Virut, Sogou, and NESIS.ay are used for training, and Neris, Menti, and Murio are used for testing. 
Because each botnet shows different characteristics, there is an overall performance degrade with Random Forest, which is affected by the dominant features of the training dataset. 
Nonetheless, VAE methods validate its reliability by showing the robust performance with the generalized dataset. 
In addition, we find that RVAE method outperforms MLP-VAE method overall based on the same features and the same size of latent variables on both datasets, as you see in Table~\ref{table4} and Table~\ref{table5}. 
It can be concluded that the botnets of network traffic flow data should be detected utilizing sequential and periodic patterns.
\begin{table}[tb]
\caption{Results comparison-Trained and tested on scenario 1,2 and 9}
\label{table4}
\begin{center}
\begin{tabular}{|c|c|c|c|c|c|c|}
\hline
Model&Recall&Precision&F1&AUPRC&AUROC\\
\hline
RVAE&0.978&0.957&0.967&0.960&0.966\\
MLP-VAE&0.974&0.959&0.966&0.959&0.966\\
Random Forest&\textbf{1.000}&\textbf{0.998}&\textbf{0.999}&\textbf{1.000}&\textbf{1.000}\\
\hline
\end{tabular}
\end{center}
\end{table}

\begin{table}[tb]
\caption{Results comparison-Trained and tested on the datasets in Table ~\ref{table3}}
\label{table5}
\begin{center}
\begin{tabular}{|c|c|c|c|c|c|c|}
\hline
Model&Recall&Precision&F1&AUPRC&AUROC\\
\hline
RVAE&\textbf{0.969}&0.892&\textbf{0.929}&\textbf{0.972}&\textbf{0.975}\\
MLP-VAE&0.944&0.891&0.917&0.967&0.962\\
Random Forest&0.424&\textbf{0.982}&0.592&0.888&0.901\\
\hline
\end{tabular}
\end{center}
\end{table}


\begin{figure}[tb]
\centering 
\subfloat[duration of window : 5s]{%
  \includegraphics[width=\linewidth]{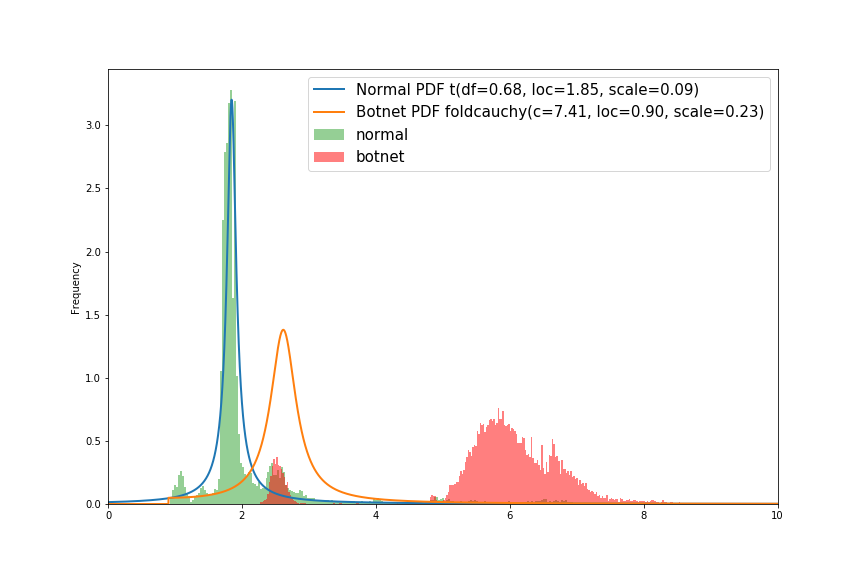}
  \label{fig:recon_error_5}
}\newline
\subfloat[duration of window : 60s]{%
  \includegraphics[width=\linewidth]{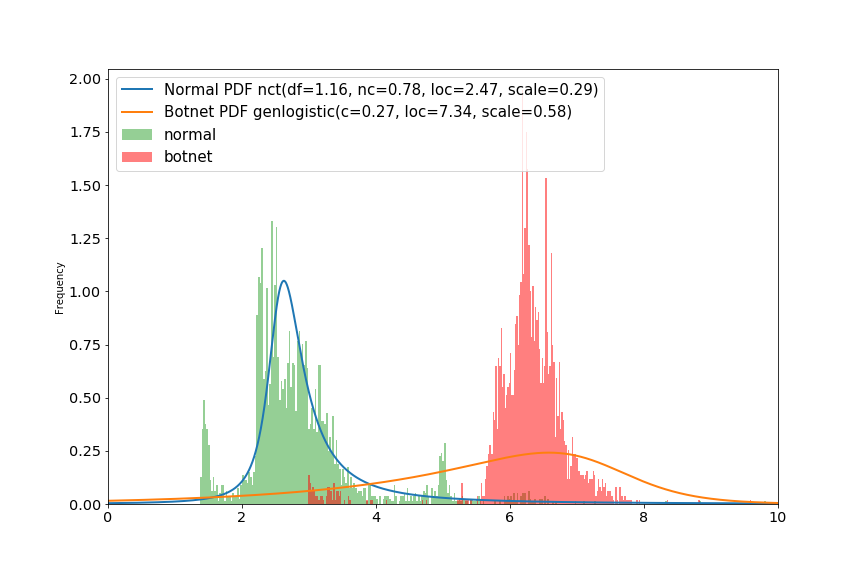}
  \label{fig:recon_error_60}
}\newline
\subfloat[duration of window : 300s]{%
  \includegraphics[width=\linewidth]{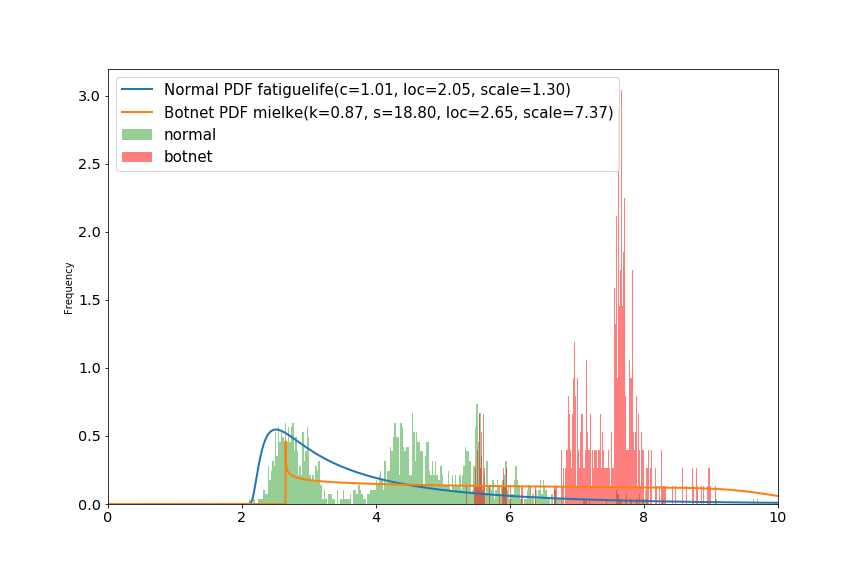}
  \label{fig:recon_error_300}
}
\caption{Distribution of reconstruction error}
\label{fig:recon}
\end{figure}

\textbf{Probability Density Function of reconstruction errors}
As shown in Fig.~\ref{fig:recon}, the distribution of the reconstruction errors of botnet samples can be distinguished from the distribution of the normal sample reconstruction errors. 
As we only use {\em non-malicious} samples for training, we expect that the reconstruction errors of {\em malicious} samples are larger than that of the  {\em non-malicious} samples. 
Comparing medians of those two distributions, we intuitively notice that the median of the distribution of {\em non-malicious} reconstruction errors is larger than the median of the distribution of botnet reconstruction errors, even if the estimated PDF function may not perfectly represent the samples in the testing dataset, as the {\em best-fit PDF} is determined with the validation dataset.  

Especially, you can find a group of botnet samples which have the smaller reconstruction errors compared to the other botnet samples in Fig.~\ref{fig:recon_error_60}.
We focus on the samples whose reconstruction errors are smaller than 4. 
We find that 66\% of the samples of the scenario 6 labeled as botnet show the reconstruction errors less than 4, while only 0.0\% -- 4.0\% of samples show reconstruction errors less than 4 in the other scenarios (1,2,8,9).
The scenario 6 utilizes proprietary command control channels unlike other scenarios most of which use IRC, HTTP and P2P communication methods~\cite{garcia2014empirical}.
The samples of the group having small reconstruction errors show low values for DNS, smtp, ssl, the number of IP addresses, the number of ports, and the number of different IP addresses in the time window. 
These characteristics mainly represent {\em non-malicious} samples other than the botnet samples.
We conclude that the general nature which can be found in the scenario 6 makes dozens of samples belonging to the scenario obtain the smaller reconstruction errors. 


\begin{table}[tb]
\caption{Results comparison}
\label{table6}
\begin{center}
\begin{tabular}{|c|c|c|c|c|c|c|}
\hline
Window&\multirow{2}{*}{Recall}&\multirow{2}{*}{Precision}&\multirow{2}{*}{F1}&\multirow{2}{*}{AUPRC}&\multirow{2}{*}{AUROC}\\
duration(s)&&&&&\\
\hline
5&\textbf{1.000}&0.865&0.928&0.791&0.881\\
60&0.969&\textbf{0.892}&\textbf{0.929}&\textbf{0.972}&\textbf{0.975}\\
300&0.998&0.537&0.699&0.905&0.972\\

\hline
\end{tabular}
\end{center}
\end{table}


\textbf{Duration of window}
In order to propose the right duration of window, our experiments have been done with changing the duration of window to 5 seconds, 60 seconds and 300 seconds, as you see in Table~\ref{table6}.
In general, the results of 60-second duration of window are higher than those of other duration lengths.
We infer that as we use a long duration, the number of source IP addresses which belongs to the same time window increases, which aggravates the vanishing gradients problem in a long-term sequence.
On the other hand, too short duration cannot provide efficient length to represent the patterns of the time windows with statistically aggregated values. 
Therefore, it is crucial to decide the appropriate duration of the time window.
From our experiments, 60-second duration is the most suitable, as you can find in Table~\ref{table6} quantitatively and Fig.~\ref{fig:recon} qualitatively.
In particular, the precision with the 300 seconds duration is low relative to the others. 
As shown in Fig.~\ref{fig:recon_error_300}, the PDF of the botnet instances contains the large part of the distribution of the {\em non-malicious} instances, which is followed by the low precision. 
Comparably, as shown in Fig.~\ref{fig:recon_error_60}, as most of the botnet samples can be represented by the {\em best-fit PDF}, the PDF of the botnet reconstruction errors is the most clearly distinct from the PDF of {\em non-malicious} reconstruction errors with the 60-second duration.


In addition, we highlight that the types of {\em best-fit PDF} vary depending on the duration of windows.
As shown in Fig.~\ref{fig:recon}, while the {\em best-fit PDF} for the botnets with the 60 seconds window is {\em generalized logistic} distribution, the best distribution for the botnets with the 5 seconds window is {\em fold cauchy}. Also, the best estimated PDF for the botnets with the 300 seconds window is {\em Mielke} distribution. 
This shows that setting the duration of window has a significant effect on the performance of the anomaly detection system.

\section{conclusion}
In this paper, we validate RVAE anomaly detection method taking into account for the sequential and periodic nature for the network traffic flow data. The study is of significance to providing the applicable solution for the botnet detection system, especially in an online manner. Moreover, as the proposed method is validated on various scenarios of botnet operation, including the botnets which are not used for training, it can be concluded that the proposed method is robust in detecting previously unseen botnets. 

For future studies, we plan to study some improvements in the proposed method. 
First, fuzzy logic can be adapted to improve the anomaly detector utilizing PDF.
It can provide more logical and systematic way of using PDFs for anomaly detection than comparing likelihoods from two distributions.
In addition, it is potential to improve performance of the anomaly detector if the method to cope with some cases of botnets having the small reconstruction errors from the normal cases is developed.
The common characteristics of the cases of botnets, which use a proprietary protocol, can be utilized to develop such a method. 
Moreover, the various VAE or RVAE architecture can be adapted to improve its anomaly detection performance.


\bibliographystyle{ieeetr}
\bibliography{main.bib}

\end{document}